\documentclass[twocolumn,showpacs,preprintnumbers,amsmath,amssymb,prd,letterpaper,floatfix,nofootinbib,superscriptaddress,]{revtex4-1}
\usepackage{graphics}
\usepackage{graphicx}
\usepackage{epsfig}
\usepackage[usenames]{color}
\usepackage{longtable}
\PassOptionsToPackage{hyphens}{url}\usepackage{hyperref}
\usepackage{breakurl}
\usepackage{latexsym}
\usepackage{amsmath}
\usepackage{amsthm}
\usepackage{amsfonts}
\usepackage{amssymb}
\usepackage{dsfont}
\usepackage{bm}

\newcommand{\E}{\mathrm{e}}  
\newcommand{\be}{\begin{equation}}
\newcommand{\ee}{\end{equation}}
\newcommand{\bea}{\begin{eqnarray}}
\newcommand{\eea}{\end{eqnarray}}

\begin{document}

\title{ $B_s\pi^+$ scattering and search for $X(5568)$ with lattice QCD}

\author{C.~B.~Lang}
\email{christian.lang@uni-graz.at}
\affiliation{Institute of Physics,  University of Graz, A--8010 Graz, Austria}

\author{Daniel Mohler}
\email{mohler@kph.uni-mainz.de}
\affiliation{Helmholtz-Institut Mainz, 55099 Mainz, Germany}
\affiliation{Johannes Gutenberg Universitat Mainz, 55099 Mainz, Germany}

\author{Sasa Prelovsek}
\email{sasa.prelovsek@ijs.si}
\affiliation{Department of Physics, University of Ljubljana, 1000 Ljubljana, Slovenia}
\affiliation{Jozef Stefan Institute, 1000 Ljubljana, Slovenia}

\date{\today}

\begin{abstract}
 We investigate $B_s\pi^+$ scattering  in $s$-wave using lattice QCD in order to
 search for an exotic resonance $X(5568)$ with flavor $\bar b s \bar d u$;
 such a state was recently reported by D0 but was not seen by LHCb.   If
 $X(5568)$  with $J^P=0^+$ exists, it can strongly decay only to $B_s\pi^+$
 and lies significantly below all other thresholds, which makes a lattice
 search for $X(5568)$ cleaner and simpler than for other exotic
 candidates. Both an elastic resonance in $B_s\pi^+$ as well as a deeply bound
 $B^+\bar K^0$ would lead to distinct signatures in the energies of lattice
 eigenstates, which are not seen in our simulation.  We therefore do not find
 a candidate for $X(5568)$  with $J^P=0^+$ in agreement with the recent LHCb result. 
 The extracted $B_s\pi^+$ scattering length is compatible with zero within the error.  
\end{abstract}

\preprint{MITP/16-074}
\preprint{HIM-2016-03}
\maketitle

\section{Introduction}

The D0 collaboration reported evidence for a relatively narrow peak in
the $B_s\pi^+$ invariant mass not far above threshold \cite{D0:2016mwd}. The
peak was attributed to a resonance $X(5568)$ with mass $m_X=5567.8\pm
2.9{+0.9\atop -1.9}~$MeV and width $\Gamma_X=21.9\pm 6.4{+5.0\atop -2.5}~$MeV
with significance 5.1$~\sigma$, while its quantum numbers were not
measured. Its decay to $B_s\pi^+$ implies exotic flavor structure $\bar b s
\bar d u$. The LHCb collaboration subsequently investigated the cross-section
as a function of  the $B_s\pi^+$ invariant mass with increased statistics and did 
not find any peak in the  same region \cite{Aaij:2016iev}. 

If the state $X(5568)$ with flavor $\bar b s \bar d u$ and some $J^P$ exists, it is unique among exotic candidates, as it can strongly decay only in one final state  $B_s\pi^+$ and is relatively far below other thresholds. This allows for a more reliable, cleaner and simpler theoretical search for it within QCD, which will be elaborated below. Most notably, the next threshold for most $J^P$ choices is $B^+\bar K^0$, which lies about 210 MeV above  $X(5568)$ and is therefore not expected to  play a notable role for this state.  The only nearby threshold is $B_s^*\pi^+$, which lies within the width of $X(5568)$ and it couples to this state only if its quantum numbers are $J^P=1^-$. 
 
Most theoretical studies which accommodate a $X(5568)$ propose $J^P=0^+$. A
number of QCD sum-rule studies do find $X(5568)$    (for example
\cite{Agaev:2016mjb,Wang:2016mee,Chen:2016mqt}), but these assume  that a  continuum of scattering
states starts above the isolated $X(5568)$ pole, which is a questionable
approach for a resonance. Reference \cite{Xiao:2016mho} finds  $X(5568)$  as a $B^+\bar K^0$ bound state, while 
all other works
\cite{Agaev:2016urs,Albuquerque:2016nlw,Burns:2016gvy,Guo:2016nhb} disfavor
this option in view of the large binding energy $\simeq 210~$MeV. A state is
also found within the tetraquark models \cite{Wang:2016tsi,Liu:2016ogz} and
quark models \cite{Stancu:2016sfd}, while other quark model studies
\cite{Lu:2016zhe,Chen:2016npt} do not confirm it. The  approaches based on
Hybridized Tetraquarks  \cite{Esposito:2016itg} and Unitarized Effective field
theory \cite{Albaladejo:2016eps,Kang:2016zmv}  do not favour its existence. A number of
physics scenarios were considered in \cite{Burns:2016gvy,Guo:2016nhb}, all
disfavouring the $X(5568)$.

In this paper we present the first study of $B_s\pi^+$ and $B^+\bar K^0$
scattering within lattice QCD in order to search for the $X(5568)$.  We
consider the channel $J^P=0^+$ where  $B_s\pi^+$ and $B^+\bar K^0$ are in
$s$-wave, which is favoured by several phenomenological studies (for example
\cite{Agaev:2016mjb,Xiao:2016mho,Stancu:2016sfd,Wang:2016tsi,Liu:2016ogz}).
The major simplification in the ab-initio lattice search for   $X(5568)$ comes
from the fact that it can strongly decay only to $B_s\pi^+$, while the next
relevant threshold $B^+\bar K^0$ is significantly higher. The task is
therefore   to study elastic  $B_s\pi^+$ scattering. During the last decade,
the lattice community has successfully demonstrated the extraction of
hadronic resonances that appear in elastic scattering (see for example
\cite{Dudek:2012xn,Pelissier:2012pi,Lang:2011mn,Prelovsek:2013ela,Mohler:2012na})
by determining the scattering matrix using the so-called L\"uscher formalism
\cite{Luscher:1990ux,Luscher:1991cf}. We apply the same well-established
formalism to determine whether  $B_s\pi^+$ scattering has a resonant or
non-resonant shape. For completeness, we consider both channels $B_s\pi^+$ and
$B^+\bar K^0$ coupled together, which should render a lattice signature for
$X(5568)$ even if it was predominantly a deeply bound $B^+\bar K^0$ state. Note that $B_s\pi^+$ scattering is elastic in the wide region below  $B^+\bar K^0$, and there the well-tested formalism is reliable.  

Section II provides an analytic prediction for the energies of lattice
eigenstates in case an $X(5568)$ claimed by D0 existed. The technical details of simulation and analysis are
elaborated in Section III. The eigenenergies from the actual
simulation are presented in Section IV, where a comparison to the analytic prediction is made.  We conclude that the results from the simulation do not support the existence of  $X(5568)$ with $J^P=0^+$.  

 \section{Expected signatures of $X(5568)$ }
 
 The lattice simulation determines the energies of QCD eigenstates with given
 quantum numbers for finite spatial size $L$. We consider the quantum numbers
 $J^P=0^+$,  the flavor content $\bar bs\bar du$, total momentum zero, while
 the spatial size of our lattice is $L\simeq 2.9~$fm.   Before presenting the
 energies obtained from the simulation, we illustrate what would be the
 distinct features in the spectrum if $X(5568)$ exists. We will argue that  an eigenstate with energy $E\simeq m_X$ is expected in a scenario with $X(5568)$, while there is no such eigenstate in absence of $X(5568)$.  
 
 \subsection{Resonance in $B_s\pi^+$}\label{sec:res}
 
\begin{figure}[tb]
\begin{center}
\includegraphics*[width=0.49\textwidth,clip]{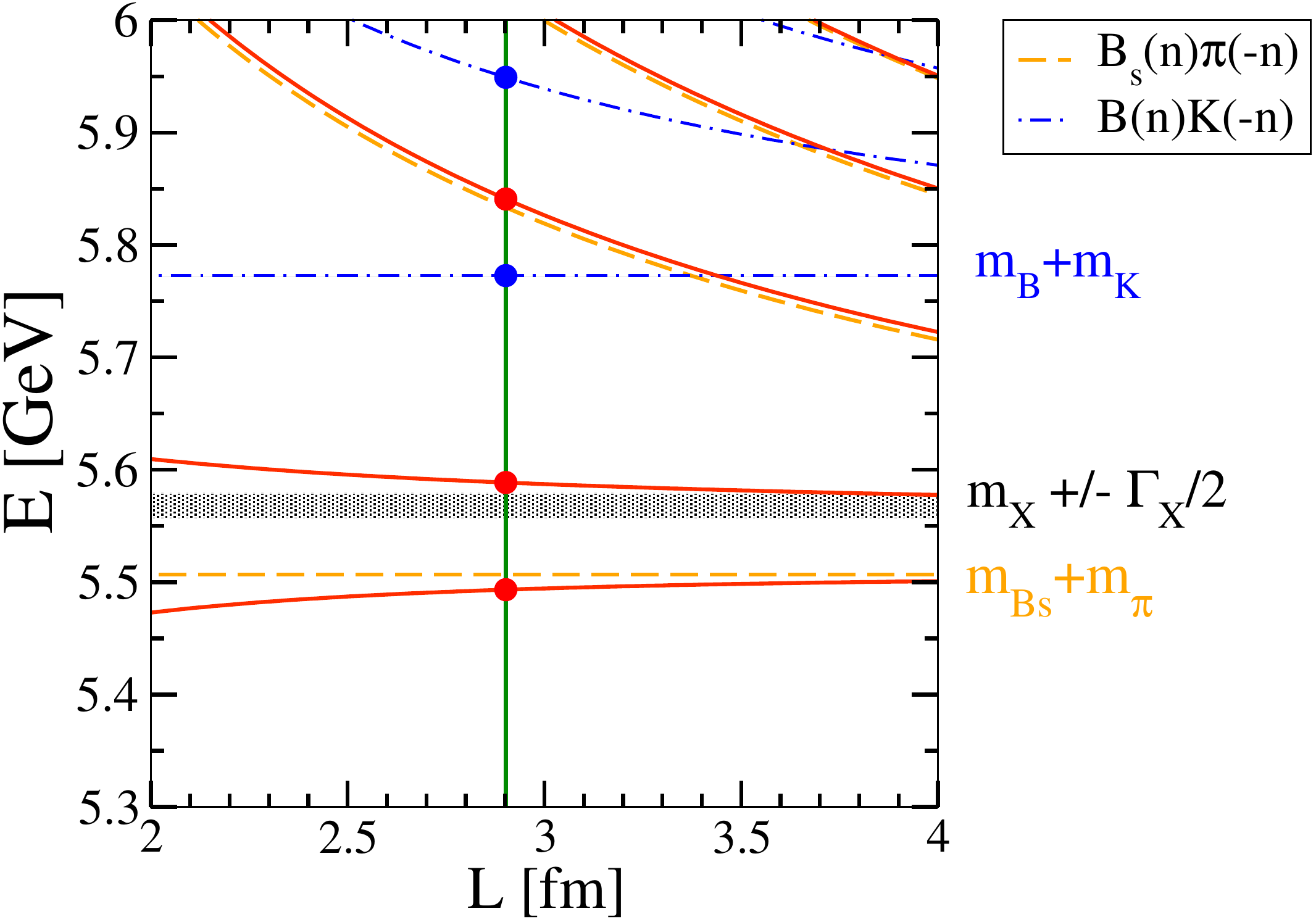}
\end{center}
\caption{Analytic predictions for energies $E(L)$ of  eigenstates as a
  function of lattice size $L$:  red solid lines are $B_s\pi$ eigenstates in
  the scenario with resonance  $X(5568)$   \cite{D0:2016mwd}; orange dashed
  lines are $B_s\pi$   eigenstates when $B_s$ and $\pi$ do not interact;
  blue dot-dashed lines are $B^+\bar K^0$ eigenstates  when $B^+$ and $\bar
  K^0$ do not interact; the grey band indicates the position of $X(5568)$ from
  the D0 experiment  \cite{D0:2016mwd}. Physical masses of hadrons are used. 
  The lattice size $L=2.9~$fm, that is used in our simulation, is marked by the vertical line. }
\label{fig:analytic}
\end{figure}

 The $X(5568)$  appears  as a peak in the $B_s\pi^+$ invariant mass 
 and is naturally considered as an  elastic resonance in $B_s\pi^+$, 
whatever the origin of this exotic state  may be.   The hypothesis with and
without a resonance lead to very distinct spectra of eigenenergies, as shown by solid and dashed lines in Figure \ref{fig:analytic}. In case $B_s$ and $\pi^+$ do not interact, they have back-to-back momenta $\mathbf{p}=2\pi \mathbf{n}/L$  due to the periodic boundary conditions in space, and the     energies of $B_s(n)\pi^+(-n)$ eigenstates (momenta in units of $2\pi/L$ are given in parentheses) 
 \begin{equation}
 E^{n.i.}(L)=\sqrt{m_{B_s}^2+\left(\frac{2\pi \mathbf{n}}{L}\right)^2}+\sqrt{m_{\pi}^2+\left(\frac{2\pi \mathbf{n}}{L}\right)^2},\;\; \mathbf{n}\in N^3
 \end{equation}
 are represented by the dashed orange lines. 
 The red solid lines represent
 the expected energies of the $B_s\pi^+$ system in case of a resonance $X(5568)$. 
 They result from the resonant Breit-Wigner-type  phase shift   
 \begin{equation}
 \label{BW}
 \delta_{B_s\pi}(p)= \mathrm{atan}\left[\frac{E\; \Gamma(E)}{m_X^2-E^2}\right] ,\quad \Gamma(E)=\Gamma_X\frac{p(E) m_X^2}{p(m_X) E^2}\,,
 \end{equation}
where $m_X$ and $\Gamma_X$ are the observed mass and width of $X(5568)$  \cite{D0:2016mwd}. 
The  (infinite-volume) elastic phase shift $\delta(E)$  and the discrete energies of  eigenstates $E$ on the lattice of size $L$ are related 
via the rigorous L\"uscher's relation \cite{Luscher:1990ux,Luscher:1991cf}
\begin{align}
 \label{luscher}
 \delta_{B_s\pi}(p)=\mathrm{atan}\biggl[\frac{\sqrt{\pi} p L}{2\,Z_{00}(1;(pL/2\pi)^2)}\biggr] ~ \\
 \mathrm{where} \ E(p)=\sqrt{m_{B_s}^2+p^2}+\sqrt{m_{\pi}^2+p^2} ~.\nonumber
  \end{align}  
  The   eigen-energies $E$ in  the scenario with $X(5568)$ are  obtained by inserting the resonant phase shift    (\ref{BW})  to the left-hand-side of (\ref{luscher}) and solving   for   discrete $p$, which gives  $E$.    The resulting discrete eigen-energies $E(L)$ are shown for a range of $L$ 
  by the red curves in    Figure \ref{fig:analytic}.     The resonant scenario  predicts an eigenstate near $E\simeq m_X$ (red solid), while there is no such eigenstate for $L=2-4~$fm in a scenario with no or small interaction between $B_s$ and $\pi^+$ (orange dashed).  These are distinct and robust features in the spectra, which do not get modified  for different   parametrisations in the resonant scenario, or for  different interaction in the non-resonant scenario.

At and above the $B^+\bar K^0$ threshold,  these states also appear as
eigenstates and will be considered in our simulation. For our lattice
parameters $X(5568)$ is far below $B^+\bar K^0$ threshold and one would not
expect a strong influence from that channel unless the dynamics leads to a really strong coupling. The dot-dashed blue lines show  the energies of non-interacting $B^+(n)\bar K^0(-n)$ in the limit when both channels are decoupled.   
  
 \subsection{Deeply bound $B^+\bar K^0$}\label{sec:BS}
 
 Next we consider the unlikely scenario where the $X(5568)$ is a very deeply
 bound  $B^+\bar K^0$ state,  in the limit where it is decoupled from
 $B_s\pi^+$. Then a simulation would render an eigenstate with $E\simeq m_X$
 up to exponentially small correction in $L$, with $\lim _{L\to \infty }
 E(L)=m_X$. In addition there would be almost non-interacting states
 $B_s(n)\pi^+(-n)$ and  $B^+(n)\bar K^0(-n)$ near orange and blue lines,
 respectively. For the simulated  $L\simeq 2.9~$fm, the number of eigenstates is therefore the same as for the resonant scenario. The values of expected energies are also similar, up to the small energy shifts. This remains true in a scenario with  a deeply bound  $B^+\bar K^0$ state which also couples to $B_s\pi^+$.

 \section{Lattice simulation details}
 
 \subsection{Gauge configurations}
  
 We employ gauge configurations from the PACS-CS collaboration, with $N_f=2+1$
 dynamical quarks, lattice spacing $a=0.0907(13)~$fm, $V=32^3\times 64$,
 $L\simeq 2.9~$fm and $m_\pi=156(7)(2)~$MeV \cite{Aoki:2008sm}. Our own fit
 for the pion mass yields a somewhat larger value of $162.6(2.2)(2.3)$MeV. The light and
 strange quarks are non-perturbatively improved Wilson fermions. 
 
 Closer inspection of that ensemble shows that there are
 a few configurations responsible for a strong fluctuation of the pion
 mass. In our analysis we consider both the full set of gauge configurations
 and a subset where four configurations \footnote{  The PACS-CS configurations leading to largest fluctionations are  hM-001460, jM-000260, jM-000840 and jM-000860. } leading to strong fluctuations in the
 pion mass are removed. This results in a pion mass for the subset roughly $6~$MeV larger than
 quoted above.   We demonstrate below that our final conclusions are independent of this
 choice.

  \subsection{Quark mass parameters}
  
For the light up/down quarks the mass parameter of the original simulation is used \cite{Aoki:2008sm}. For the
 strange quark we use a partially quenched  setup with the valence mass $m_{s}^{val}$ closer to the physical point than the dynamical sea quark mass $m_{s}^{dyn}$
 \cite{Aoki:2008sm}, leading to  $m_K=504(1)(7)$~MeV
 \cite{Lang:2014yfa}. 
 The bottom quark is treated as a valence quark using the Fermilab method
 \cite{ElKhadra:1996mp,Oktay:2008ex},  where the kinetic masses ($M_2$) are
 tuned to experiment and the energy differences  are less prone to
 discretization effects compared to the energies themselves. The bottom quark mass is fixed as discussed
in   \cite{Lang:2015hza} which renders a spin-averaged  kinetic mass   $\tfrac{1}{4}(M_2^{B_s}+3M_2^{B_s^*})=5.086(135)(73) ~$GeV somewhat smaller than in experiment $E^{exp}_{\overline{B_s}}=\tfrac{1}{4}(m_{B_s} + 3m_{B_s^*} ) = 5.4032(18)~$GeV. The mass splittings of various hadrons containing a $b$-quark are in good agreement with experiment (see Table II of  \cite{Lang:2015hza}).
     
\subsection{Dispersion relations}\label{subsec:disprel}

Here we discuss the dispersion relation between energy, mass and momentum of
the pion and $B_s$. This is needed  to determine the  $s$-wave scattering
length $a_0$ for $B_s\pi$ scattering from the ground state lattice energy
$E^{lat}_{gr}$   using L\"uscher's relation
\cite{Luscher:1990ux,Luscher:1991cf}. The  momentum $p_{gr}$  is obtained from  $E_{gr}^{lat}=E_\pi(p_{gr})+E_{B_s}(p_{gr})$. For the pion we use the relativistic dispersion relation
\be
E_\pi(p)=(m_\pi^{lat}+p^2)^{1/2}
\ee
and for the heavy meson $B_s$ the Fermilab dispersion relation \cite{ElKhadra:1996mp,Oktay:2008ex}
\be
E_{B_s}(p)=M_1+p^2/(2M_2)-p^4/(8M_4^3)\;.
\ee
The values $M_1=1.61246(54)$, $M_2=2.298(70)$ and $M_4=1.59(54)$ have been
determined in \cite{Lang:2015hza} by measuring  $E_{B_s}(p)$ for
 several small values of $p$  on our lattice. These are the values for the
 correlated fits with all gauge configurations.  When excluding close to
 exceptional configurations we redo the whole analysis with the reduced set.
 
Within the Fermilab approach, the rest masses have large discretization effects but mass differences are expected to be close to physical \cite{Kronfeld:2000ck} and can be compared to experiment. In order to compare the splitting $E^{lat}-\bar m^{lat}$  with  $E^{exp}-\bar m^{exp}$, we will sometimes plot
\begin{equation}
E=E^{lat}_n-E_{\overline{B_{s}}}^{lat}+E^{exp}_{\overline{B_{s}}}\,
\end{equation}
where $E_{\overline{B_{s}}}^{lat}$ is the spin-averaged ground
state energy of the $B_s$ system from our simulation and
$E^{exp}_{\overline{B_{s}}}$ is the corresponding physical energy $\tfrac{1}{4}(m_{B_s} + 3m_{B_s^*} )$ 
from experiment. 

\subsection{Lattice operators}

  To determine the energies of a system with $J^P=0^+$ and total momentum zero, we employ six interpolating fields\footnote{The interpolators transform according to the $A_1^+$ irreducible representation of discrete group $O_h$.} of meson-meson type, where each  meson is   projected to a  definite momentum: 
 \begin{align}
 \label{ops}
O_{1,2}^{B_s(0)\pi(0)}&=\left[\,\bar{b}\Gamma_{1,2} s\,\right](\mathbf{p}=0)\left[\,\bar{d}\Gamma_{1,2} u\,\right](\mathbf{p}=0)\\ 
O_{1,2}^{B_s(1)\pi(-1)}&=\!\!\!\!\!\!\!\!\!\sum_{\mathbf{p}=\pm\mathbf{e_{x,y,z}}~2\pi/L}\!\!\!\!\!\!\! \left[\bar{b}\Gamma_{1,2} s\right](\mathbf{p})\left[\bar{d}\Gamma_{1,2} u\right](-\mathbf{p})\nonumber\\ 
O_{1,2}^{B(0)K(0)}&=\left[\,\bar{b}\Gamma_{1,2} u\,\right](\mathbf{p}=0)\left[\,\bar{d}\Gamma_{1,2} s\,\right](\mathbf{p}=0)\nonumber
\end{align} 
 with $\Gamma_1=\gamma_5$ and  $\Gamma_2=\gamma_5 \gamma_t$. 
 
 One could use also local or quasi-local diquark-antidiquark operators, for example $[\bar b C\gamma_5 \bar d]_{3_c}  [s C\gamma_5 u]_{\bar 3_c}$, but these can be expressed via Fierz transformations   as $\sum_iM_1^i(p)M_2^i(-p)$ , where $M_1^iM_2^i=B_s\pi, B_s^* \rho, B_{s1}a_1, BK,B^*K^*,B_1K_1,...$ (see  \cite{Padmanath:2015era} for a detailed discussion). 
 
 The $B_s\pi$ and $BK$ are  the essential ones for the energy region near  $X(5568)$ and are already included in our choice (\ref{ops}). It remains to be seen if structures with significantly separated diquark and antidiquark   \cite{Brodsky:2014xia} could be also be probed\footnote{ In view of this, we note that the conclusions of our previous studies of  X(3872), Y(4140) \cite{Padmanath:2015era} and $Z_c^+$ \cite{Prelovsek:2014swa} apply to (quasi) local $[qq][\bar q\bar q]$, but they  do not apply to the case of significantly separated $[qq]$ and $ [\bar q\bar q]$.} by meson-meson operators like (\ref{ops}), or if specific implementation of those is needed.

\begin{figure*}[t]
\begin{center}
\includegraphics*[width=\textwidth,clip]{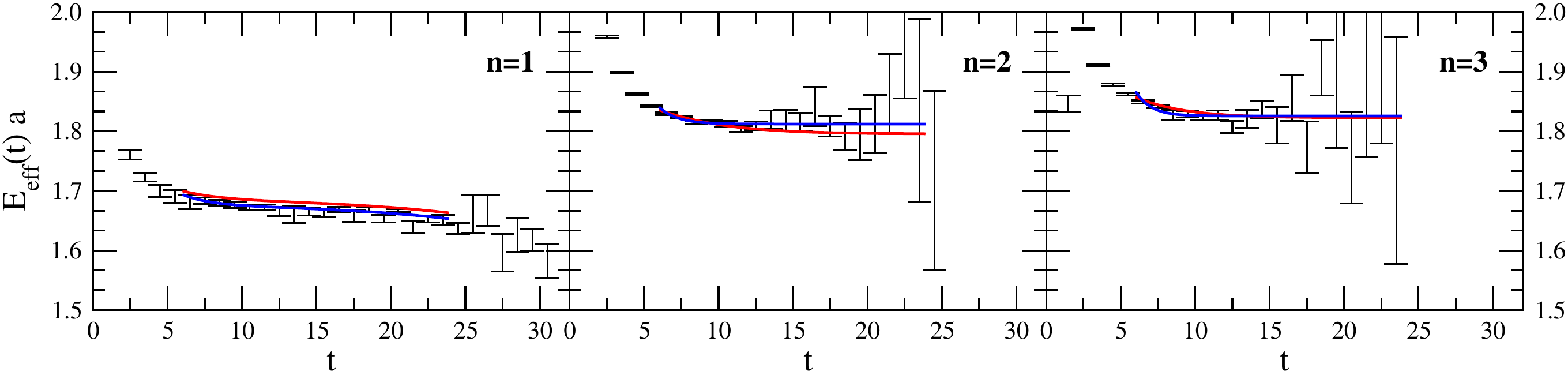}
\includegraphics*[width=\textwidth,clip]{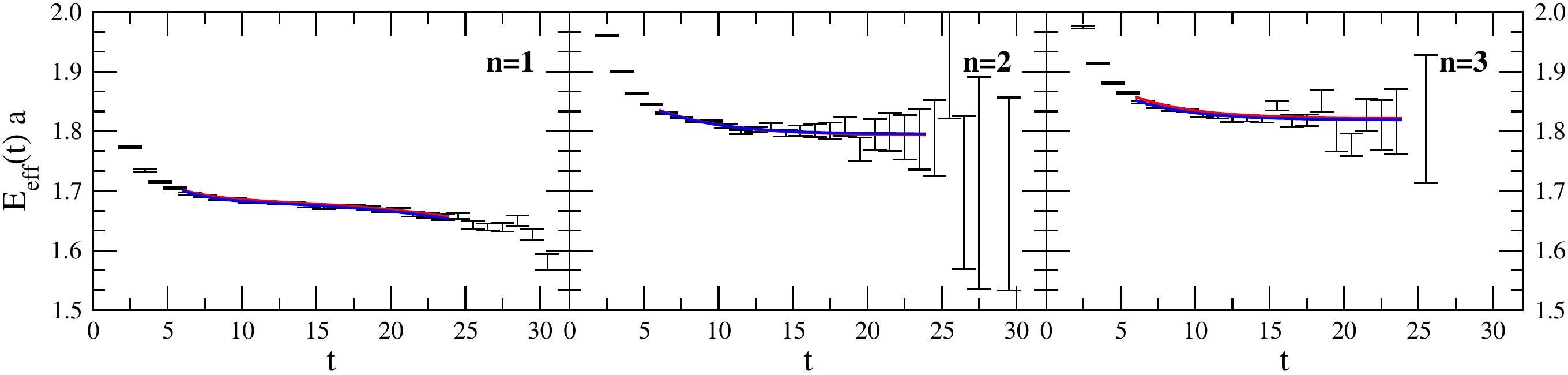}
\end{center}
\caption{ The effective energies $E_{eff}(t)\equiv\log (\lambda(t)/\lambda(t+1))$ and  
the curves due to our eigenvalue fits
over the fit range $6\le t \le 24$  for the cases ``all,A'' (top panes) and
``all-4,A'' (bottom panes)  defined in the caption of Figure  \ref{fig:vary}. From left
to right the plots show the ground state, first excited state and second
excited state. The fit model (\ref{eq:model_b}) was used for the ground state, while the form (\ref{eq:model_a})  was employed for
the higher states. The correlated (uncorrelated) fits are shown by red (blue) curves.}
\label{fig:meff}
\end{figure*}

\subsection{Correlation matrix and eigenenergies}

The energies  $E_n$ of eigenstates $|n\rangle$ are obtained from the 
correlation matrix 
\begin{align}
\label{C}
C_{jk}(t)&= \langle \Omega| O_j (t^\prime +t) O_k^\dagger (t^\prime)|\Omega \rangle
=\sum_nZ_j^nZ_k^{n*}~e^{-E_n t}~
\end{align}
which also contains the information on the overlaps $Z_j^n\equiv \langle \Omega|{\cal O}_j|n\rangle$.  
 
All quark lines run between source and sink, there are no ``backtracking'' loops. There are only two
diagram types: (a)  $(B_s\leftrightarrow B_s)(\pi\leftrightarrow\pi)$ and 
$(B\leftrightarrow B)(K\leftrightarrow K)$ and (b) $B_s\pi\leftrightarrow BK$ 
where the $s$ and $u$ quarks exchange partners. The Wick contraction gives for (a) a product of two
traces, for (b) only one trace.

The correlation matrix elements are calculated using the stochastic distillation method
proposed in \cite{Morningstar:2011ka}. In the distillation  method \cite{Peardon:2009gh}
the quark fields in the interpolators are smeared according to  
$q\equiv \sum_{k=1}^{N_v}v^{(k)}v^{(k)\dagger}q_{point}$; in the
stochastic version \cite{Morningstar:2011ka} one uses  random combination of the sources. 
We use $N_v=192$ eigenvectors of the lattice laplacian 
$v^{(k)}$ reducing them to 16 combinations.
The method is convenient  for calculating a variety of Wick contractions. The details of our implementaion are  presented in  \cite{Lang:2014yfa} where we apply it to $D_s$ states.

Energies $E_n$ and overlaps $Z_j^n$ are extracted from the correlation matrix $C_{jk}(t)$ using  the generalized
eigenvalue method \cite{Michael:1985ne,Luscher:1985dn,Luscher:1990ck,Blossier:2009kd}
\begin{align} 
 C(t)u^{(n)}(t)&=\lambda^{(n)}(t)C(t_0)u^{(n)}(t)~,
\end{align}
where $\lambda^{(n)}(t)\propto  e^{-E_n t}$ at large $t$.  Correlated and uncorrelated fits to $\lambda^{(n)}(t)$ are used and $t_0=2$.  

\subsection{Choice of operator subsets}

Although we compute the full $6\times 6$ correlation matrix we attempt to
minimize the statistical noise by choosing subsets of most important operators. The guiding principle is the 
stability of the overlap factors $Z_n^i(t)$ over the fit range and the statistical noise 
of the eigenvalues. From the overlap factors we identify the dominantly contributing
lattice operators to each eigenstate.

\subsection{Energy fits}

Ideally the eigenvalues follow a pure exponential behaviour. Due to the limited set of
operators there are contaminating contributions from higher excitations at small
propagation distances. The finite-time effects (like backward propagation) due
to the anti-periodic boundary conditions in time and $n_T=64$ are
important for large distances and even more for light particles. For this reason one chooses a fit model that, in addition to the leading exponential
form, allow  for such contributions.

For larger energies and propagation distances much less than $n_T/2$ the finite size effects are
negligible and we fit $\lambda(t)$ to
\be\label{eq:model_a}
f(t)=a_1 \,\E^{-E_1\,t} +a_2 \,\E^{-E_2\,t}
\ee
ensuring that $E_1 < E_2$. The second term effectively represents possible higher
excitation visible at small $t$ values, allowing for a larger fit range. We use this form
for eigenstates above the ground state.
 
 For two-meson eigenstates with
light particles and long propagation time one has to choose a form that can represent
also (a) the propagation back in time, (b) the propagation of one meson in one direction and the
second meson in the opposite direction of time (see, e.g., Appendix in \cite{Prelovsek:2010kg}).
In our study this concerns the $B_s\pi$ ground state and we use
\begin{eqnarray}\label{eq:model_b}
f(t)&=&a_1 \left(\E^{-E_1\,t} +\E^{-E_1\,(n_T-t)} \right)\\
&&+a_2 \left(\E^{-E_2\,t} +\E^{-E_2\,(n_T-t)} \right)\nonumber \\
&&+a_3  \left(\E^{-m_\pi t- m_{B_s}(n_T-t)}+\E^{-m_{B_s} t- m_\pi(n_T-t)}\right)\nonumber
\end{eqnarray}
(again checking that $E_1 < E_2$) and where $m_\pi$ and $m_{B_s}$ have been
determined from the corresponding single-meson correlators (see Subsection \ref{subsec:disprel}).

Figure \ref{fig:meff} gives an example of these fits showing the effective energies $E_{eff}(t)\equiv \log (\lambda(t)/\lambda(t+1))$ and  our fits, where fit range is determined based on $\chi^2/d.f.$.   The errors-bars of the final energy values correspond to statistical errors obtained using single-elimination jack-knife. 

\subsection{Scattering length}\label{sec:scattlength}

We determine the  $s$-wave scattering length $a_0$ for $B_s\pi$ scattering from the ground state energy $E^{lat}_{gr}$ on the lattice using L\"uscher's relation \cite{Luscher:1990ux,Luscher:1991cf}
 \begin{equation}
 a_0^{B_s\pi}\equiv \lim_{p\to 0} \frac{1}{p\cot\delta(p)}=\frac{\sqrt{\pi}L}{2Z_{00}(1;(p_{gr}L/2\pi)^2)}
 \end{equation}
The  momentum $p_{gr}$  is obtained as discussed earlier in Subsection \ref{subsec:disprel}. 

\section{Results}\label{sec:results}
 
\begin{figure}[t]
\begin{center}
\includegraphics*[width=0.38\textwidth,clip]{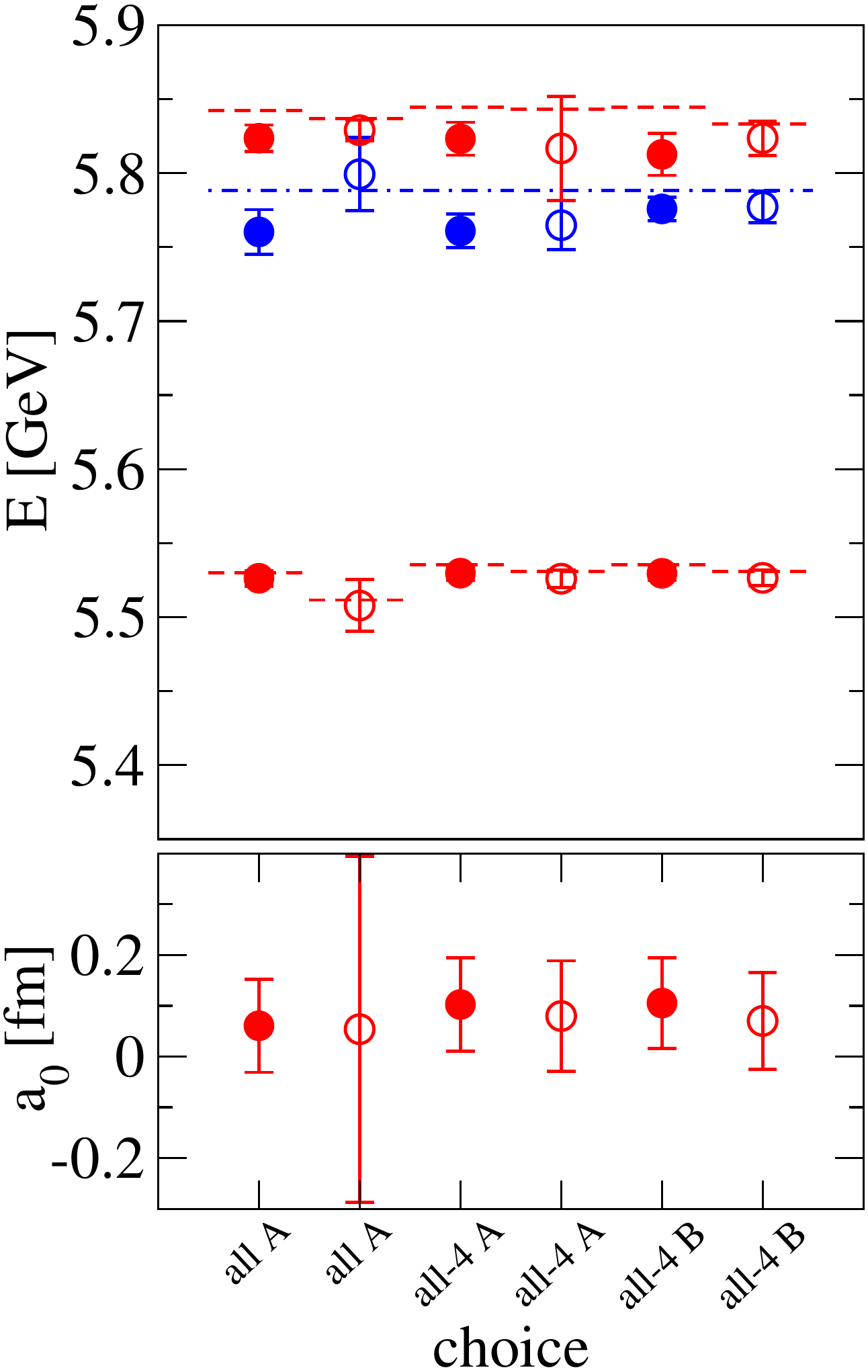}
\end{center}
\caption{The eigenenergies of  the $\bar bs\bar du$ system with $J^P=0^+$
  from a lattice simulation for various choices. The horizontal lines show energies of eigenstates $B_s(0)\pi^+(0)$, $B^+(0)\bar K^0(0)$ and $B_s(1)\pi^+(-1)$ in absence of interactions.  
The energies $E=E^{lat}_n-E_{\overline{B_s}}^{lat}+E^{exp}_{\overline{B_s}}$
are shown, where the spin-averaged ground state is set to its physical
value. The sets with full symbols are from correlated fits to the $t$-dependence for
$B_s\pi$ states while open symbols result from uncorrelated fits. Notation ``all''
refers to the full set of gauge configurations while ``all-4'' refers to the set with four
(close to exceptional) gauge configurations removed. Set A is from 
interpolator basis $O_{1}^{B_s(0)\pi(0)},O_{1}^{B_s(1)\pi(-1)},O_{1}^{B(0)K(0)}$  while set B results from a larger basis  $O_{1}^{B_s(0)\pi(0)},O_{1,2}^{B_s(1)\pi(-1)},O_{1,2}^{B(0)K(0)}$.}
\label{fig:vary}
\end{figure}

We aim to determine the energies of eigenstates for the system with flavor $\bar b s \bar d u$, $J^P=0^+$ and total momentum zero, and compare them to analytic predictions for the scenarios in the previous section.

As mentioned above, we present the final energies as
$E=E^{lat}_n-E_{\overline{B_{s}}}^{lat}+E^{exp}_{\overline{B_{s}}}$. 
The upper pane of Figure \ref{fig:vary} shows the results for the energy levels from correlated
(full symbols) and uncorrelated (open symbols) fits to the $t$-dependence for the
three lowest eigenstates for two choices of interpolator basis (A and B) and two set of gauge
configurations (``all'' and ``all-4''). While there is a visible difference between
those choices for single energy levels, the extracted value for the $B_s\pi$  scattering
length displayed in the bottom pane of Figure \ref{fig:vary} is largely
independent of these choices. Furthermore none of these variations lead to an
energy level in close vicinity to the X(5568).

\begin{figure}[tb]
\begin{center}
\includegraphics[width=0.49\textwidth,clip]{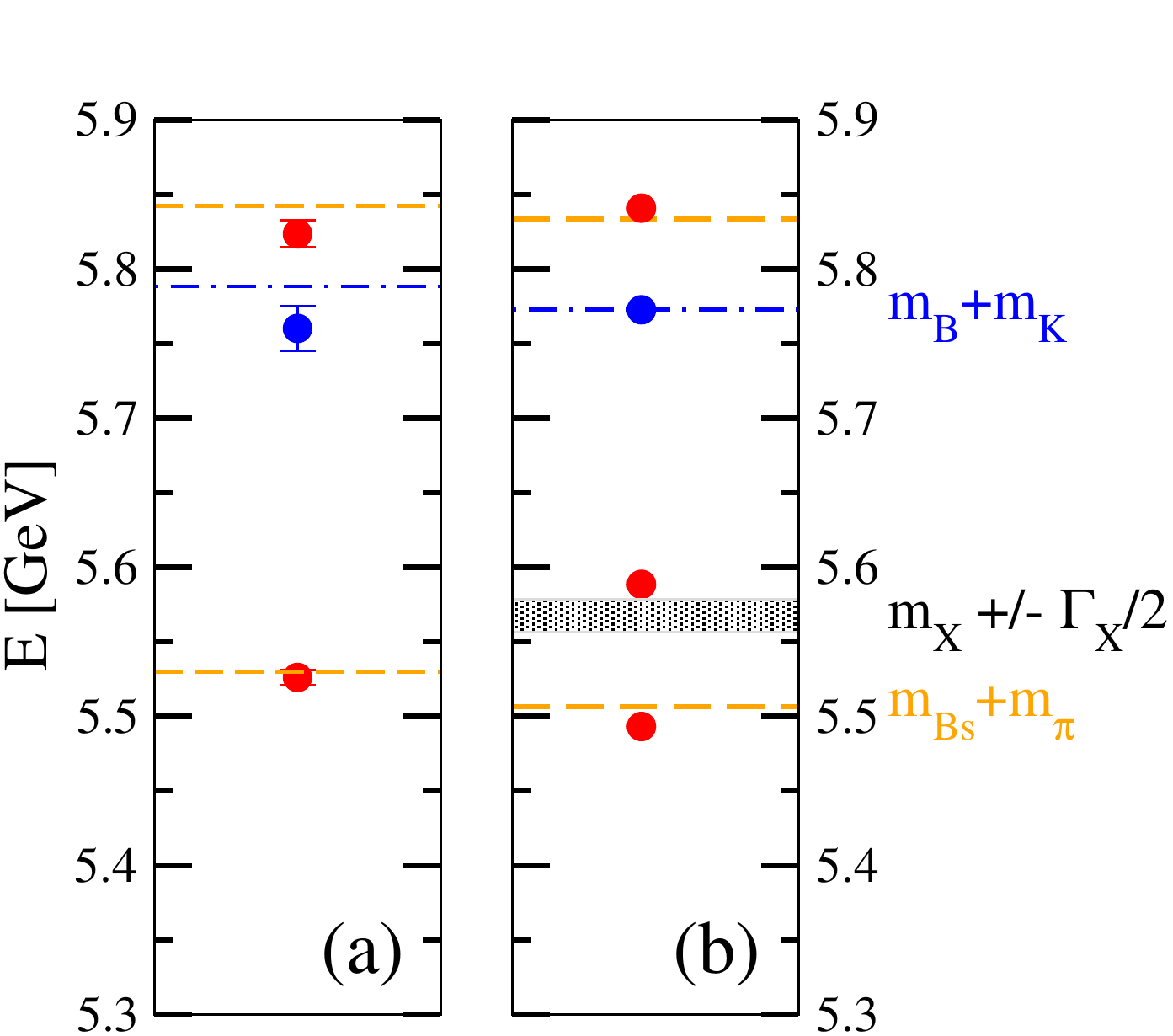}
\end{center}
\caption{(a) The eigenenergies of  the $\bar bs\bar du$ system with $J^P=0^+$
  from our lattice simulation    and  (b) an  analytic prediction based on
  $X(5568)$, both at lattice size $L=2.9~$fm.  The horizontal lines show energies of eigenstates $B_s(0)\pi^+(0)$, $B^+(0)\bar K^0(0)$ and $B_s(1)\pi^+(-1)$ in absence of interactions; momenta in units of $2\pi/L$ are given in parenthesis.  
The pane (a) shows the energies
$E=E^{lat}_n-E_{\overline{B_s}}^{lat}+E^{exp}_{\overline{B_s}}$ with the
spin-averaged $B_s$ ground state set to its experiment value.  The pane (b) is based on the
experimental mass of the $X(5568)$ \cite{D0:2016mwd}, given by the grey band,  and experimental masses other particles. }
\label{fig:results}
\end{figure}

Our final results for the eigenenergies of the $\bar bs\bar du$ system with $J^P=0^+$
obtained from our simulation  are presented  in  Figure
\ref{fig:results}a  (correlated fit, choice ``all A'' from Figure \ref{fig:vary}). The circles in Figure \ref{fig:results}b  show the
analytic predictions for the spectrum at our $L=2.9~$fm if a resonance
$X(5568)$ exists (same as in Figure \ref{fig:analytic}).       The analytic
prediction based on  $X(5568)$ renders an energy level near $E\simeq m_X\sim
5.57~$GeV, which is not observed in the actual simulation (left figure).
Results of our simulation therefore do not provide support for the existence
of an $X(5568)$ resonance with $J^P=0^+$ in $B_s\pi$ scattering. The scenario
of $X(5568)$ as a deeply bound $B^+\bar K^0$ also would render an energy level near  $E\simeq m_X$, so this scenario is also not supported by the simulation.

The spectrum from the simulation is closer to the non-interacting limit
indicated by the horizontal lines in Figure  \ref{fig:results}, indicating a
rather weak interaction.    These lines show energies $m_{B_s}+m_{\pi}$,
$m_{B}+m_{K}$ and $E_{B_s(1)}+E_{\pi(-1)}$ of the relevant two-meson states
$B_s(0)\pi(0)$, $B^+(0)\bar K^0(0)$ and $B_s(1)\pi(-1)$ in absence of
interactions: in pane (a) the lines show the sum of single-particle energies  obtained
from the simulation, while pane (b) is based on the experimental masses.
The lowest and highest  eigenstates $|n\rangle$   have large overlap $\langle O|n\rangle$  with the  $O^{B_s\pi}$ interpolator  (red) and the middle state has large overlap with the $O^{BK}$ interpolator (blue),  which confirms their identification.

The resulting $B_s\pi$  scattering length $a_0$ (cf. Section \ref{sec:scattlength}) is small and compatible with zero within errors, as displayed in Figure \ref{fig:vary}. Our result is  compatible with $a_0^{D_s\pi}=-0.002(1)~$fm obtained  for a similar channel  $D_s\pi$  
 from their ground eigenstate \cite{Liu:2012zya}. Using the value from
 reference  \cite{Liu:2012zya,Liu:2008rza} as an input, the  Chiral perturbation theory (ChPT) \cite{Liu:2009uz} for $B_s\pi$ and Unitarized ChPT \cite{Altenbuchinger:2013vwa} for $D_s\pi$ also leads to a very small scattering length in
agreement with our lattice result.

 \section{Conclusions}
 
 If  the exotic state $X(5568)\to B_s\pi^+$ observed by D0 exists, it could be
 one of the easiest exotic candidates to establish on the lattice.  The state
 $X(5568)$ with the most natural quantum number $J^P=0^+$ would represent a
 resonance in elastic $B_s\pi^+$ scattering, significantly below the next
 relevant threshold $B^+\bar K^0$.   We presented the first simulation of $B_s\pi^+$ scattering in the channel $J^P=0^+$, aiming to  search  for  possible exotic resonances close to the threshold. For completeness we  took into account also the $B^+\bar K^0$ channel, which has a threshold $210~$MeV above $X(5568)$. 
 In a system with a resonance, L\"uscher's formalism predicts an eigenstate with  $E\simeq m_X$ if $X(5568)$ exists, while such an eigenstate is not found in our simulation. Our results therefore do not support the existence of  $X(5568)$ with $J^P=0^+$. Instead, the results appear closer to the limit where $B_s$ and $\pi$ do not interact significantly.  \\
 
  \vspace{0.3cm}
  
 {\it Node added:} After this manuscript  appeared as preprint, the analytic study \cite{Lu:2016kxm} presented  the finite-volume spectrum in this channel based on the Unitarized ChPT \cite{Altenbuchinger:2013vwa}.   
  Their analytic conclusion agrees with  our conclusion from the lattice simulation. 
 \vspace{0.5cm}
 
 {\bf Acknowledgments}
 
 \vspace{0.2cm}
  
We thank the PACS-CS collaboration for providing the gauge
configurations. S.P. thanks R. Lebed and M.  Pappagallo for clarifications
on the experimental result. D.M. would like to thank M. Hansen for useful
discussions. This work is supported in part by the  Slovenian Research Agency
ARRS and by the Austrian Science Fund project FWF:I1313-N27. The calculations
were performed on Fermilab USQCD clusters and on computing clusters at  the
University of Graz (NAWI Graz). We thank the Fermilab Lattice Collaboration
for allowing us to calculate the observables at Fermilab using previously stored propagators.


\end{document}